\documentclass[aps,pra,nofootinbib,preprint,amsmath,amssymb,floatfix]{revtex4-2}
\usepackage{color}
\usepackage{graphicx}

\begin{document}

\newcommand{\tc}{\textcolor}
\newcommand{\g}{blue}
\newcommand{\ve}{\varepsilon}
\title{ Possible connection  between time-splitting parameter and surface tension in Casimir-type  problems
 \emph{}}         

\author{ Iver Brevik }      

\affiliation{Department of Energy and Process Engineering, Norwegian University of Science and Technology, N-7491 Trondheim, Norway}

\date{\today}          

\begin{abstract}
We discuss  four different, though related, fundamental topics related to the Casimir effect: 1) We suggest that the application  of  Casimir theory to real dielectric materials, thus implying the atomic spacing as a course-grained length parameter, makes it natural to assume that this parameter is of the same order of magnitude  as  the QFT  time-splitting parameter multiplied with the velocity of light. 2) We show that application of  Casimir theory to a thick fluid shell (apparently a closed mechanical system), leads actually to  an unstable situation if not extra mechanical forces, typically surface tension forces,  are brought into consideration. 3) We analyze how the presence of a radial Casimir repulsive pressure modifies the filling process of a spherical vacuum hole in an infinite fluid (the Reynolds problem), with the result that a bounce occurs at a finite though very small  radius. 4) As a comment on an apparently similar situation in general relativity, we consider the gravitational collapse of a singular shell. It might seem natural to allow for the presence of a repulsive Casimir pressure in this case also, thereby obtaining a bounce-like situation again. However, we have to conclude that such a procedure implies an omission of  the Casimir field's gravitational energy, and is therefore hardly tenable, although it is in our opinion worth mentioning.

\end{abstract}
\maketitle

\section{Introduction. Analysis of a dielectric ball}

From a fundamental viewpoint, delicate issues appear when the quantum field theory (QFT) of the Casimir effect is applied to a dielectric condensed body.  As QFT is basically a theory derived for vacuum surroundings,  there is no length scale involved (looking away from the extreme Planck scale of $10^{-33}~$cm). Related to the absence of such a scale, there appear infinities in the  theory, which are regularized in various ways.  One often makes use of  a time-splitting parameter $ \tau$, which after a Wick rotation implies a splitting in imaginary time.  It is natural to ask:  is there a physical significance of this parameter? As we will argue, it is possible to suggest that there is  a physical connection  between $\tau$ and the interatomic spacing of a condensed dielectric medium.

Consider in this context, as a very typical example, the Casimir theory of a nonmagnetic dielectric ball of radius $a$. This theory  was  worked out by Milton back in 1980 \cite{milton80} by using the time-splitting method  (cf. also the related exposition in Ref.~\cite{hoye17}).  We  confine ourselves here to give only the essentials. The material in the ball is assumed nonmagnetic and is for simplicity also assumed to be dilute, $\varepsilon-1 \ll 1$, where $\varepsilon$ is the permittivity. The surroundings are vacuum. The  stress $P$ on the surface $r=a$ is calculated as the difference between the $rr$ electromagnetic stress tensor components on the outside and the inside. Some calculation gives the expression
\begin{equation}
P= - \frac{(\varepsilon-1)^2\hbar c}{16^2\pi a^4 }\left[ \frac{16}{\delta^3}+ \frac{1}{4} \right], \quad \delta= \frac{\tau c}{a}. \label{1}
\end{equation}
Both terms are negative, corresponding to an attractive force. We shall be interested in the cutoff dependent part only. Calling this $P_{\rm cutoff}$, we have
\begin{equation}
P_{\rm cutoff}= -\frac{(\varepsilon-1)^2}{16\pi }\frac{\hbar}{ac^2}\frac{1}{\tau^3}. \label{2}
\end{equation}
A striking property of this expression is that the parameter $\tau$ (dimension seconds)  becomes naturally related to the surface tension coefficient $\sigma$  (dimension N/m), the latter known from ordinary  hydrodynamics.

The following estimate is illustrative: Forget about Casimir effect for a while, and start from the expression (\ref{2}) as it stands. Require that this inward pressure is equal to the inward surface pressure following from Laplace's formula, i.e.,
\begin{equation}
\frac{(\varepsilon-1)^2}{16\pi }\frac{\hbar}{ac^2}\frac{1}{\tau^3}= \frac{2\sigma}{a}. \label{3}
\end{equation}
This estimate gives us information about how $\tau$ is related to ordinary physical quantities. First, we observe the connection
 \begin{equation}
\tau \propto \sigma^{-1/3}, \label{4}
\end{equation}
which is in this special case   independent of the radius $a$ (it is not a general property).  We may here insert $\sigma = 0.073~$N/m as for a water-air surface, and  choose a reasonable value $\varepsilon-1= 0.01$ for the diluteness parameter. For the quantity  $\tau c$, the path travelled by light during the time $\tau$, we thus get
\begin{equation}
\tau c \sim 1~\rm{ \AA}, \quad \tau \sim 10^{-19}~ \rm{s}. \label{4a}
\end{equation}
This is the point we wished to emphasize:  the minimum distance  $\tau c$ corresponds to typical {\it interatomic distances}. Is this merely a numerical coincidence? It might be so, but we think it is worth noticing that the time-splitting parameter $\tau$ coming from QFT becomes so naturally linked to the natural 'lattice spacing' in condensed matter physics.

\section{On the stability of a closed mechanical system under the influence of Casimir forces}

We go on with a related example, also that apparently simple, but nevertheless being of fundamental interest. Let us consider the
 force balance for a closed (an insulated) mechanical system when subjected to Casimir forces. Adopt the same fluid sphere as above, but now with a spherical hole of radius $r=a$ in it. The outer radius is renamed as $r=b$. The fluid body is accordingly an annular spherical region bounded by the  surfaces $a$ and $b$. For $r<a$ and $r>b$ there is a vacuum.

We will calculate the Casimir force on the two spherical surfaces. To circumvent problematic counter terms, we will assume that the fluid medium has a permittivity $\varepsilon$ and a permeability $\mu$ such that
  \begin{equation}
  \varepsilon \mu =1.\label{5}
  \end{equation}
  For this kind of medium, called an isorefractive medium  (cf., for instance, Refs.~\cite{brevik82,brevik83,ellingsen09}) the refractive index is unity. As a slightly more complicated situation we may take the inner medium 1 to have material constants $\varepsilon_1, \mu_1$ and outer medium 2 to have material constants $\varepsilon_2, \mu_2$, both cases satisfying the condition (\ref{5}).

  From Ref.~\cite{brevik82} we have for the Casimir pressure on the surface $r=b$,
\begin{equation}
P(b)= \frac{0.09235}{8\pi}
\frac{\hbar c}{b^4}\left( \frac{\mu_{12}-1}{\mu_{12}+1}\right)^2 \left[ 1+0.311\frac{\mu_{12}}{(\mu_{12}+1)^2}\right], \label{6}
\end{equation}
where  $\mu_{12} \equiv \mu_1/\mu_2 = \mu$.   An important property of this expression  is that it is invariant under the substitution $\mu_{12} \rightarrow \mu_{21}$. We may thus apply the expression (\ref{6}) to find the pressure $P(a)$ on the inner  surface $r=a$ also, only by substituting $b$ with $  a.$ Both pressures are outward directed.

Consider a narrow cone corresponding to a solid angle $d\Omega$, so that a surface $a^2d\Omega$ is crossed out at $r=a$ and  $b^2d\Omega$ is crossed out at $r=b$. The Casimir radial force on the fluid within the cone, assuming for simplicity the medium to be dilute, is
\begin{equation}
F_{\rm Casimir}= \frac{0.09235\hbar c}{32\pi}(\mu-1)^2\left( \frac{1}{a^2}+\frac{1}{b^2}\right)d\Omega.\label{7}
\end{equation}
When summing up  the forces acting on the actual fluid body, we have in principle to consider the electrostrictive force also. This force will however be without influence here, since  it is a gradient force. To see this, it  is sufficient to give the expression for the force density in the usual case of a nonmagnetic medium,
\begin{equation}
{\bf f}_{\rm elstr} = \frac{1}{2}\varepsilon_0 {\bf \nabla}
\left( E^2\rho \frac{d\varepsilon}{d\rho}\right). \label{8}
\end{equation}
As written, this expression assumes a nonpolar medium, so that $\varepsilon$ is a function of $\rho$ only. The derivative $d\varepsilon /d\rho$ may  be calculated via  the Clausius-Mossotti relation.

The electrostrictive   force  is  outward directed  at $r=a$ and  inward directed  at $r=b$. These two opposing forces press the  fluid together, but the total radial force $F_{\rm elstr}$ on the actual fluid body is zero, in view of Gauss' law. (In order to detect the electrostrictive effect, one has in general to do local pressure experiments.)

In fluid mechanics it is physically legitimate to assume ideal conditions, so that there is no viscosity to counteract fluid motion. We thus become enfaced with a paradoxical situation: the Casimir force should  in principle be able to accelerate the fluid outwards. The sphere should fly apart. There must accordingly be   extra forces, not included so far, that make the system stable. Quite evidently, it is the surface tension that  is the remedy here. The surface pressure is  $2\sigma/r, \, r=a,b$ on the two surfaces, both pressures acting inward. Multiplying with the appropriate surface areas, it follows that the total radial force on the considered volume from surface stresses is $2\sigma(a+b)d\Omega$.

Let us require  the two  kinds of forces to balance each other out, thus
 \begin{equation}
 \frac{0.09235\hbar c}{32\pi}(\mu-1)^2\left( \frac{1}{a^2}+\frac{1}{b^2}\right) = 2\sigma (a+b). \label{9}
\end{equation}
When $a \ll b$ we get approximately
\begin{equation}
\frac{\hbar c}{640\pi}(\mu-1)^2\frac{1}{a^2b}=\sigma. \label{10}
\end{equation}
Take $\sigma = 0.073$ N/m as before, and take   $|\mu-1| = 0.01, \, b/a= 10$.
Remarkably enough, this leads to results of the same order of magnitude as already given above, in Eq.~(\ref{4a}). Realistic input values for the surface tension as well as for the other parameters give results in rough agreement with condensed matter atomic dimensions.

The Casimir force thus cannot stand 'alone' when acting on a closed dielectric system; it needs the cooperation with conventional effects in fluid mechanics to make the situation physically consistent. Note that the {\it closeness} of the system is an essential point here. The usual situation with two parallel plates exposed to an attractive force is different as it is an {\it open} system: an external force is needed to hold the plates in place.

\section{On the filling of a spherical hole in a fluid, in the presence of the  Casimir force}

We now enter a class of  dynamical problems for which the Casimir force comes into play.  In this section we consider an infinite isotropic dielectric fluid of density $\rho$, with a central hole of radius $r=a$ in it. If the pressure at infinity is $p_\infty$, there will be a force aiming to fill the hole. For times $t<0$ we assume the situation is static, via the help of external stabilizing agencies.  At $t=0$ these agencies are removed and the hole starts filling. We regard this problem as a Newtonian one, in the sense that  relativistic effects are omitted. The fluid is  assumed to be nonviscous and incompressible so that the velocity of sound is infinite.

For simplicity we assume the fluid to be isorefractive such as before, so that it  satisfies the condition (\ref{5}). Thus within the hole $\varepsilon_1=\mu_1= 1$, while on the outside $\varepsilon_2= 1/\mu_2$.  In view of the already noted invariance of the expression (\ref{6}) under the interchange $\mu_{12} \rightarrow \mu_{21}$, we can use this expression to calculate the surface pressure $P(a)$  in the initial static period $t<0$. Moreover, we can use the same expression to calculate the pressure $P(R(t))$ at later times $t>0$ also, where $R(t)$ is the instantaneous radius of the hole (note that $R(0)=a$). The physical reason for this is that the characteristic time for establishment of electromagnetic equilibrium,  roughly of order $R(t)/c$,   is much smaller than the characteristic time scale for the fluid motion.

We now write the surface pressure in the form
\begin{equation}
 P(R(t))=    \frac{C}{8\pi R^4(t)},  \label{11}
 \end{equation}
 where $C$ is a constant. The Euler equation for the velocity $v=v(t)$ at an arbitrary point in the fluid is
 \begin{equation}
 \frac{\partial v}{\partial t}+v\frac{\partial v}{\partial r}= -\frac{1}{\rho}\frac{\partial p}{\partial r}, \label{12}
 \end{equation}
 where $p=p(r,t)$ is the pressure. Continuity of the radial flow implies that
 \begin{equation}
 r^2v(r,t)=F(t), \label{13}
 \end{equation}
 where $F(t)$ is an unspecified function (this can also be seen from the incompressibility condition $\nabla^2\phi=0$, with $\phi$ the velocity potential). Thus we can write
 \begin{equation}
 \frac{\dot{F}(t)}{r^2}+\frac{\partial}{\partial r}\left( \frac{1}{2}v^2\right) = -\frac{1}{\rho}\frac{\partial p}{\partial r}. \label{14}
 \end{equation}
 For a fixed value of $t$ we integrate this equation from   $R(t)$  to infinity,
 \begin{equation}
 -\frac{\dot{F}(t)}{R(t)} +\frac{1}{2}V^2(t)= \frac{p_\infty}{\rho}-\frac{P(R(t))}{\rho}, \label{15}
 \end{equation}
 where $V(t)= \dot{R}(t)$ is the velocity of the surface. It is here assumed that the velocity at infinity is equal to zero.

 We now apply Eq.~(\ref{13}) at the instantaneous surface,
 \begin{equation}
 R^2(t)V(t) = F(t). \label{16}
 \end{equation}
 so that in view of  $d/dt = Vd/dR$ we can rewrite the first term in (\ref{15}) as
 \begin{equation}
 -\frac{\dot{F}}{R}= -2V^2-\frac{1}{2}R\frac{dV^2}{dR}, \label{17}
 \end{equation}
 so that (\ref{15}) reduces to
 \begin{equation}
 \frac{dV^2}{dR}+\frac{3V^2}{R}= \frac{C}{4\pi \rho R^5}-\frac{2p_\infty}{\rho R}. \label{18}
 \end{equation}
With an integrating factor  $\exp {\int (3/R)dR} = R^3$ we can integrate this equation analytically,
\begin{equation}
V^2= \frac{a^3}{R^3}\left[ \frac{2p_\infty}{3\rho}\left( 1-\frac{R^3}{a^3}\right) -\frac{C}{4\pi \rho}\frac{1}{Ra^3}\left( 1-\frac{R}{a}\right) \right]. \label{19}
\end{equation}
where we have made use of the initial condition $V(0)=0$.

Taking the square root of this expression, using $ V= dR/dt$, we can by integration over $R$ follow the time dependence of the filling process. For our purpose it is however of greater interest to analyze the surface velocity in the limit of small  $R/a$. We see that  the velocity becomes zero at a radius $R=R_{\rm min}$, where
\begin{equation}
R_{\rm min}= \frac{3C}{8\pi p_\infty a^3}. \label{20}
\end{equation}
The collapse of the fluid bounces off  at this radius. Quantitatively, from  (\ref{6}) and (\ref{11}) we get
\begin{equation}
C= 0.09235\hbar cf(\mu_{12}) = 2.9\times 10^{-27}f(\mu_{12}), \label{21}
\end{equation}
where $f(\mu_{12})$ is the material-dependent function in (\ref{6}). If we approximate this function by unity, and choose as input values $p_\infty = 1~$mPa, $a= 1~$mm, we find
\begin{equation}
R_{\rm min}  \approx 3~\rm{\AA}. \label{22}
\end{equation}
At this value the hydrodynamical description of the fluid as a continuous medium is however questionable.  The sensitivity of $R_{\rm min}$ with respect to $a$ is very pronounced. If we had inserted for instance $a= 1~$cm, $R_{\rm min}$ would have been very low and the continuum model would definitely have been lost.

Let us summarize the physics of this process: the filling of the hole is initiated by the pressure $p_\infty$, and the time dependence of $R(t)$ is further dependent on the constants $\rho$ and $\mu_2$. The very existence of the bounce is however due to the Casimir repulsion, i.e., a quantum mechanical effect.

In the absence of the Casimir modification, the solution to the above problem can be found in Landau-Lifshitz \cite{landau}. It was originally solved by  Rayleigh in 1917.

\section{Gravitational collapse of a singular shell, in the presence of  the Casimir force}

As  an analogy to the example in the previous section, we will now consider the gravitational collapse of a singular  spherical  shell when account is taken of the repulsive Casimir force. A shell  is a very simple material system but nevertheless  able to demonstrate  main features of the collapse process when the  Casimir surface pressure is brought into play, in a very simple way..

Let the shell whose gravitational (Schwarzschild) ) mass is $M$, start off the collapse at zero velocity when the  radius is infinity (we now put $c$, as well as Newton's gravitational constant $G$, equal to 1). The collapsing shell surface separates spacetime into two regions with different extrinsic curvature  tensors $K_{ij}$. The time-dependent shell radius is $R=R(\tau)$, where  $\tau$ is proper time.

 The external geometry is Schwarzschild,
 \begin{equation}
 ds^2= - \left( 1-\frac{2M}{R}\right)dt^2+ \left( 1-\frac{2M}{R}\right)^{-1}dr^2+r^2d\Omega^2,
  \end{equation}
  while the inner geometry is flat,
\begin{equation}
ds^2= -dt^2+ dr^2+ r^2d\Omega^2.  \label{23}
\end{equation}
In this section we let  $\sigma$ be the proper surface mass energy, and $u^\alpha$ the four-velocity of the matter. The surface energy-momentum tensor can the be written as
\begin{equation}
S^{\alpha\beta} = \sigma u^\alpha u^\beta,
\end{equation}
 where $S^{\alpha\beta}$ is a result of integrating the local interior energy-momentum tensor in the shell, $T^{\alpha\beta}$, in the normal direction $\bf n$ across the shell,
\begin{equation}
S^{\alpha\beta}= \int_{-\eta}^{\eta}T^{\alpha\beta}dn, \label{24}
\end{equation}
with $\eta =0^+$.

From the equation of motion $ {{S_i}^m}_{|m}=0$ (see below) we obtain $(\sigma R^2)_{,\tau}=0$, so that $4\pi R^2\sigma$ is a constant. In view of our initial condition $\dot{R}=0$ at $R=\infty$, the constant becomes equal to the Schwarzschild mass \cite{lightman75},
\begin{equation}
4\pi R^2\sigma = M.
\end{equation}
The four-acceleration for particles on the shell is \cite{lightman75,israel66,israel67,misner73}
\begin{equation}
D{\bf u}/d\tau= u^iu^j_{|i}{\bf e}_j +K_{ij}u^iu^j{\bf n},\label{25}
\end{equation}
where
$u^j_{|i}$ is the covariant derivative with respect to the three-geometry of the hypersurface swept out by the two-dimensional shell. The basis vectors of this surface are ${\bf e}_i= (\partial/\partial \tau, \partial/\partial \theta,\partial/\partial \varphi)$. The orthogonality relations are
\begin{equation}
{\bf e}_i\cdot {\bf e}_j=g_{ij}, \quad {\bf e}_i\cdot
{\bf n}=0,
 \quad {\bf n\cdot n}=1. \label{26}
\end{equation}
We will look for  the projection of the four-acceleration $\bf a$ onto $\bf n$. From Eq.~(\ref{25}),
\begin{equation}
{\bf a}= K_{ij}u^iu^j {\bf n}. \label{27}
\end{equation}
From the equation of motion one can derive \cite{lightman75}
\begin{equation}
{\bf a}^+ + {\bf a}^- =0, \label{27a}
\end{equation}
which generalizes the equation ${\bf a}=0$ for a free particle, and also
\begin{equation}
{\bf a}^+ - {\bf a}^-  =4\pi \sigma {\bf n}.
\end{equation}
This is the conventional theory for the collapse. Can this theory be generalized, in the sense that we allow for a repulsive Casimir force on the spherical surface in the same way as we did in the previous section? Let us start from  the pressure $P$ in the same form as before,
\begin{equation}
P= \frac{C}{8\pi R^4}, \label{28}
\end{equation}
where now $C$ has the value corresponding to a singular shell \cite{milton78},
\begin{equation}
C= 0.09235 \hbar c, \label{29}
\end{equation}
in physical units. It might seem possible to modify the equation of motion by multiplying the mean acceleration $ ({\bf a}^+ + {\bf  a}^-)/2$ by $\sigma$ and put it equal to the Casimir pressure,
\begin{equation}
\frac{1}{2}\sigma ({\bf a}^+ + {\bf a}^-) = \frac{C}{8\pi R^4} {\bf n}. \label{30}
\end{equation}
Such a procedure was actually followed in Ref.~\cite{brevik84}, leading to the following equation of motion,
\begin{equation}
(1+\dot{R}^2)^{1/2}= 1-\frac{C}{2MR}+\frac{M}{2R}\frac{1}{1-C/2MR}, \label{33}
\end{equation}
which satisfies ${\dot R}=0$ at $R=\infty$.

The presence of the factor $C$ means that the collapse bounces off at a minimum radius $R=R_{\rm min}$, where
\begin{equation}
R_{\rm min}= \frac{C}{2M}. \label{35}
\end{equation}
At first sight, this result is striking and points towards  the existence of a gravitational analogy to the bouncing effect discussed in the previous section. The argument, consisting in the use of Eq.~(\ref{30})
 has however the drawback that it neglects the gravitational influence from the Casimir energy. We therefore cannot  regard the  formalism  as physically satisfactory after all. We confine us here  only to mentioning this  point without going into further detail. If we put $C=0$ in Eq.~(\ref{33}), we get
 \begin{equation}
 (1+\dot{R}^2)^{1/2}= 1+\frac{M}{2R},
 \end{equation}
  which implies that the collapse occurs down to $R=0$.

  There are several papers discussing the importance of Casimir energy in a gravitational context under various circumstances; cf., for instance, Refs.~\cite{milton14,coronell21},  and in the present context, Ref.~\cite{vick86}.

\section{Summary}

By means of some examples, we have in this paper  elucidated how the application of the Casimir force to ordinary dielectric matter can be more delicate than usually assumed. Let us  summarize:

\noindent 1. The QFT formalism, although not containing any length scale in itself (Planck scale omitted), will when applied to a condensed medium be confronted with a natural length, namely the interatomic spacing. In Sec.~I we pointed out a possible physical interpretation of the conventional  mathematical time-splitting parameter $\tau$  as being of the same order as the interatomic transit time. A  numerical estimate supported this suggestion.

\noindent 2. The thick fluid shell model considered in Sec.~II,  at first sight apparently a stable mechanical system (gravity omitted), needs under influence from the Casimir force assistance from extra forces  to secure stability. The natural choice of such a local force is the surface tension. For simplicity, we assumed  the medium to be of the so-called isorefractive kind. A numerical estimate with realistic input parameters gave the same order of magnitude result as in Sec.~I.

\noindent 3. In Sec.~III we analyzed the filling of a vacuum hole in an infinite fluid (the Reynolds problem), when taking into account the radial repulsive Casimir force, with the result that the collapse becomes halted. There occurs a Casimir-induced bounce at a finite small radius $R_{\rm min}$. The bigger the initial radius $a$ of the hole, the smaller value is obtained for  $R_{\rm min}$.

\noindent 4. In Sec.~IV we commented briefly on an analogous situation from general relativity, namely  the collapse of a singular spherical shell. A simple application of the standard repulsive Casimir pressure to the gravitational equation of motion (\ref{30}) led to an equation of motion for the radius $R$ that implied a bounce, thus in apparent qualitative agreement with the collapse considered in Sec.~III. However, as such a procedure ignores the gravitational influence of the Casimir energy, we have to doubt its physical reality.

\section*{Acknowledgment}

I acknowledge the very valuable comments of K. A. Milton on preliminary versions of this manuscript.



\end{document}